\documentclass[11pt]{article}
\usepackage[utf8]{inputenc}
\usepackage[preprint]{acl}
\usepackage{booktabs}
\usepackage{multirow}
\usepackage{graphicx}
\usepackage{times}
\usepackage{latexsym}
\usepackage{amsmath,amssymb,amsfonts}
\usepackage[T1]{fontenc}
\usepackage{makecell,tabularx,array}
\usepackage{enumitem}
\usepackage[utf8]{inputenc}
\usepackage{microtype}
\usepackage{inconsolata}
\usepackage{graphicx}
\usepackage{booktabs}
\usepackage{array}
\usepackage[most]{tcolorbox}
\tcbuselibrary{breakable,listings}
\usepackage{listings}
\usepackage{enumitem}
\setlist[itemize]{nosep,leftmargin=*,topsep=1pt}

\usepackage{pdflscape}
\usepackage{longtable}
\usepackage{booktabs}
\usepackage{tabularx}
\usepackage{array}
\usepackage{makecell}
\usepackage{booktabs}
\usepackage{array}
\usepackage{tabularx}
\usepackage{titletoc}   
\usepackage{tcolorbox}  
\tcbuselibrary{skins, breakable, listings} 
\usepackage{longtable}  
\usepackage{booktabs}   
\usepackage{makecell}   
\usepackage{array}      
\usepackage{geometry}   
\usepackage{tabularx}
\newcolumntype{Y}{>{\raggedright\arraybackslash}X}

\usepackage{caption}

\usepackage{algorithm}
\usepackage{algpseudocode}

\usepackage{titletoc}

\title{DemMA: Dementia Multi-Turn Dialogue Agent with Expert-Guided Reasoning and Action Simulation}

\author{
\textbf{Yutong Song}$^{1}$,
\textbf{Jiang Wu}$^{2}$,
\textbf{Kazi Sharif}$^{3}$
\textbf{Honghui Xu}$^{3}$,
\textbf{Nikil Dutt}$^{1}$,
\textbf{Amir Rahmani}$^{1}$
\\[0.5em]
$^{1}${University of California, Irvine},
$^{2}$Independent Researcher,
$^{3}$Kennesaw State University
}

\begin{document}
 \maketitle

\begin{abstract}


Simulating dementia patients with large language models (LLMs) is challenging due to the need to jointly model cognitive impairment, emotional dynamics, and nonverbal behaviors over long conversations. We present DemMA, an expert-guided dementia dialogue agent for high-fidelity multi-turn patient simulation. DemMA constructs clinically grounded dementia personas by integrating pathology information, personality traits, and subtype-specific memory-status personas informed by clinical experts. To move beyond text-only simulation, DemMA explicitly models nonverbal behaviors, including motion, facial expressions, and vocal cues. We further introduce a Chain-of-Thought distillation framework that trains a single LLM to jointly generate reasoning traces, patient utterances, and aligned behavioral actions within one forward pass, enabling efficient deployment without multi-agent inference. Extensive evaluations with experts, medical students, and LLM judges demonstrate that DemMA significantly outperforms strong baselines across multiple metrics. 

\end{abstract}

\begingroup\def\thefootnote{}\footnotetext{The code is available at \url{https://anonymous.4open.science/r/DemMA-0A4B/}}\endgroup
\section{Introduction}
In dementia research and caregiver training, the scarcity of high-quality interaction data remains a structural bottleneck~\cite{livingston2020dementia}. Due to strict privacy and ethical barriers, collecting sensitive patient data is heavily restricted~\cite{jackson2019ethics}. Consequently, public corpora, especially those capturing multimodal resources such as facial expressions and vocal prosody are virtually absent. This “data desert” is particularly acute at the level of pathological granularity: even for common types such as Alzheimer’s disease (AD), public corpora are very limited, and over 20 clinically subtypes remain largely unrepresented~\cite{gorno2011classification}. As a result, patient simulation often relies on scripted materials that fail to capture the heterogeneity of real-world care interactions.
The emergence of LLM-based generative agents offers a promising avenue for data synthesis but using generic dialogue agent models as dementia simulators is unreliable~\cite{wei2022chain,agentbench2023}.
First, without clinically grounded alignment, such models risk generating content that lacks medical rigor or may even yield unsafe guidance.
Second, in extended interactions, simulations often suffer from persona drift, whereby the model gradually regresses toward a fluent, polite, and generic assistant style~\cite{shumailov2023curse}. This “over-perfect” articulation suppresses essential markers of cognitive decline.
Consequently, although the synthesized dialogues may appear linguistically fluent, they lack patient persona fidelity, rendering them inadequate for agent training.

A core challenge in dementia simulation is to model interpretable, reproducible, and subtype-specific pathological behaviors, rather than injecting stochastic variation into model outputs~\cite{Jack2018}. Although general symptoms such as confusion and linguistic errors are shared across conditions, different dementia subtypes exhibit distinct cognitive impairments. High-fidelity simulation therefore requires mechanisms that differentiate pathological inconsistency from generic model hallucination.

In addition, dementia communication is inherently multichannel, encompassing language, affect, and behavior~\cite{bourgeois2016nonverbal}. As linguistic ability declines, clinically salient signals increasingly manifest through non-verbal expressions, which are not preserved by text-only LLMs. This limitation motivates structured textual representations that encode latent non-verbal cues to compensate for degraded speech.

Finally, from a systems perspective, long-horizon dialogue coherence is commonly handled via multi-agent architectures, which introduce substantial overhead and latency, making them unsuitable for real-time dementia care training and simulation.





%

Based on these motivations, we propose \textbf{DemMA}, a framework that fine tunes open source models for end to end dementia virtual patient generation without relying on sensitive real world data. Moreover, to address the limitation that text only simulation often underspecifies language impairments, DemMA adopts a dual track modeling paradigm:

\noindent\textbf{Intrinsic Cognitive Level:} The model explicitly captures fine-grained differences across dementia subtypes and stages, ensuring that generated incoherence reflects specific persona patterns rather than random noise.
    
\noindent\textbf{Extrinsic Expression Level:} We introduce Action Labels as a compensatory mechanism to map multimodal behaviors (e.g., motion, facial expressions, sound) into text-based labels, serving three key roles: (i) explicitly revealing latent emotions and intentions that cannot be directly expressed through ambiguous language; (ii) projecting multimodal behaviors onto a text-based interaction interface; and (iii) providing key signals to distinguish between different subtypes and disease stages. 

Our contributions are summarized as follows:

\begin{itemize}
     \item We propose a clinically grounded modeling approach within DemMA for precise representation of \textbf{Dementia Patient Personas}.
     
    \item We are the first work to introduce explicit motion, facial expression and sound labels into LLM-based generative agents. 
 
    \item We introduce a multi-agent pipeline for producing high-quality interaction dialogue data and release \textbf{DemMA-Dialogue}, the first synthetic dementia dialogue dataset covering main dementia subtypes with expert validation.
    
    \item We develop a distillation training for agent \textbf{DemMA} to internalize long-horizon planning signals, supporting low latency inference with coherent, persona-consistent interactions.
\end{itemize}

\section{DemMA}
\begin{figure*}[t]
    \centering
    \includegraphics[width=0.9\textwidth]{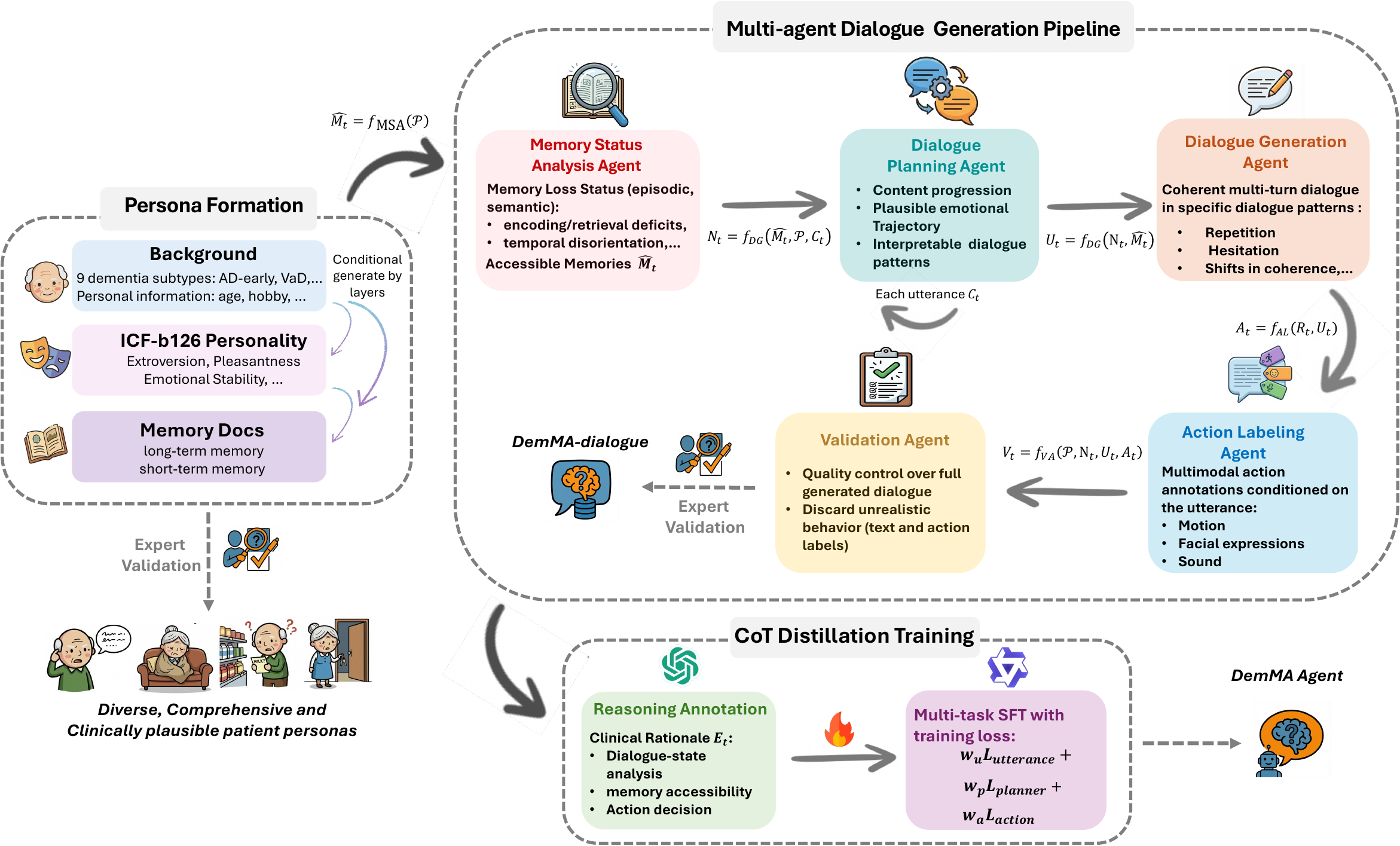} 
    \vspace{-0.5em}
    \caption{DemMA integrates three components: a) Clinically grounded patient persona formation module; b) Multi-agent dialogue dataset generation pipeline encompassing memory analysis, planning, language generation, and action simulation; and c) CoT distillation multi-task training workflow for DemMA agent.}
    \label{fig:demma_framework}
    \vspace{-1em}
\end{figure*}

Architecturally, DemMA integrates three components: (i) Clinically grounded patient persona formation module; (ii) Multi-agent dialogue dataset generation pipeline encompassing memory analysis and search, dialogue planning, language and action simulation; and (iii) CoT distillation multi-task training for DemMA agent, internalizing medical knowledge while reducing multi-agent inference latency. The overview of DemMA is shown in Fig.~\ref{fig:demma_framework}.


\paragraph{Dementia Persona Formation}
A core challenge in dementia persona simulation is enforcing longitudinal consistency and clinical validity. Naïve single-pass generation often lacks medical grounding, resulting in inconsistent profiles and persona drift. DemMA addresses this by constructing the patient persona via staged generation with explicit dependencies, propagating upstream constraints to all downstream components:
\begin{equation}
\begin{aligned}
B &\sim f_B(\cdot), \quad S \sim f_S(B),\\
M &\sim f_M(B, S), \quad \mathcal{P} \triangleq \langle B, S, M \rangle,
\end{aligned}
\label{eq:patient-formation}
\end{equation}
where $B$, $S$, and $M$ denote the personal background, personality, and memory layers, respectively, and $\mathcal{P}$ represents the resulting patient persona profile.

\paragraph{Background Layer ($B$)}
This layer encodes (i) stable demographic and life-context attributes (e.g., age, education, and comorbidities) and (ii) the dementia subtype persona along with its core clinical patterns~\cite{convery2019clinical}. We instantiate nine subtypes covering the entire spectrum: \textit{AD-early, AD-mid/late, VaD, DLB, PDD, FTD-bv, nfvPPA, svPPA, and lvPPA} (see details in Appendix~A, Table~\ref{tab:appendix-pathology-personas}). These attributes, including subtype-specific behavioral characteristics and neuropathological rationales (Appendix~A, Table~\ref{tab:appendix-persona-clinical-refs}), establish the foundational medical and contextual constraints for subsequent modeling.

\paragraph{Personality Layer ($S$)}
$S$ models interpersonal style within the WHO International Classification of Functioning, Disability and Health (ICF) \texttt{b126} psychological-function space~\cite{eerola2015alaraajaproteesia} (Appendix~\ref{sec:appendix-dementia-formation}, Table~\ref{tab:appendix-personality-icf}). Crucially, dementia-induced personality alterations are not stochastic but track the underlying neuropathology. Therefore, rather than treating personality as a disease-independent attribute, we model it as subtype-specific tendencies encoded in the ICF space (Appendix~A; Tables~\ref{tab:appendix-persona-clinical-refs} and~\ref{tab:appendix-personality-icf-dims}). This structural dependency ensures the generation of diverse yet clinically grounded interaction styles, preventing medically implausible profiles.

\paragraph{Memory Layer ($M$)} We instantiate $M$ from $(B, S)$ as (i) long-term memory (e.g., childhood experiences, life history, and salient life events) and (ii) short-term daily/weekly records of recent events, constrained to remain consistent with both long-term memory and the clinical facts in $B$ (e.g., comorbidities). This layer will characterize distinct retention and degradation patterns of remote and recent information.

\section{Multi-Agent LLM Workflow}
DemMA employs a multi-agent LLM workflow to generate long-horizon dialogue with structured control and interpretability. Separating reasoning from generation has been shown to improve coherence and reduce hallucination in complex language tasks \cite{yao2022react}. 
The five agents operate as described below.
\subsection{Memory Status Analysis Agent}
The workflow begins with the Memory Status Analysis Agent. Based on the dementia persona and corresponding memory-status information derived from clinical diagnostic literature (Appendix~\ref{sec:appendix-dementia-formation}, Table~\ref{tab:appendix-memory-status-templates}), the agent identifies accessible memory systems during the dialogue and infers clinically grounded features, such as the availability of recent, remote, and semantic memory, as well as cue responsiveness:

\begin{equation}
\hat{M}_t = f_{\text{MSA}}(\mathcal{P}),
\end{equation}
where $\hat{M}_t$ denotes the accessible memories and clinically grounded memory features at turn $t$ and $f_{\text{MSA}}$ is the memory analysis function. This inferred memory state conditions all downstream reasoning and generation. 
\subsection{Dialogue Planning Agent}
\textit The \textit{Dialogue Planning Agent} then produces a high-level plan
conditioned on the patient’s personality, inferred memory status, and dialogue context $C_t$. The agent organizes the overall conversational flow by specifying planned 
content progression and a plausible emotional trajectory across turns as:
\begin{equation}
N_t = f_{\text{DR}}(\hat{M}, \mathcal{P}, C_t),
\end{equation}
Externalizing this planning step improves transparency and control, ensuring that 
the generated dialogue follows interpretable patterns consistent with the patient’s 
cognitive and emotional state rather than opaque model behavior.

\subsection{Dialogue Generation Agent}



\textit{Dialogue Generation Agent} produces the natural-language dialogue, converting high-level plans into turn-level utterances while maintaining consistency with the patient’s cognitive limitations, personality traits, and emotional state. By explicitly conditioning generation on structured plans, the agent produces coherent multi-turn dialogue while preserving dementia-specific communication patterns such as repetition, hesitation, misplaced references, and shifts in coherence. Utterance generation at turn $t$ is modeled as:
\begin{equation}
U_t = f_{\text{DG}}(N_t, \hat{M}_t).
\end{equation}

\subsection{Action Labeling Agent}
The \textit{Action Labeling Agent} augments each dialogue turn with multimodal action annotations conditioned on the generated utterance and the inferred reasoning state. It assigns labels corresponding to plausible \textbf{Motion}, \textbf{Facial expressions}, and \textbf{Sound}, capturing behavioral signals commonly observed in dementia interactions such as fidgeting, gaze aversion, trembling speech, and sudden emotional shifts \cite{agentbench2023}. The full list of action labels is provided in Appendix~\ref{app:action-labels}.

 These action sequences provide a multimodal view of patient responses and support downstream tasks such as embodied simulation and multimodal data generation. multimodal grounding action labels predictions at turn $t$ are modeled as:
\begin{equation}
A_t = f_{\text{AL}}(R_t, U_t).
\end{equation}
\subsection{Validation Agent}

The final component is the \textit{Validation Agent}, which performs quality control over the generated dialogue. It outputs a validation score that determines whether the turn should be accepted or regenerated, reducing drift, incoherence, and unrealistic behavior in long-horizon generation. The validation process is defined as:
\begin{equation}
V_t = f_{\text{VA}}(\mathcal{P}, R_t, U_t, A_t),
\end{equation}
where $V_t$ denotes the validation score at turn $t$ and $f_{\text{VA}}$ is the validation function. A turn is accepted when: $V_t \ge \tau$,
and regenerated otherwise.







\subsection{Expert Validation}
Before finalizing the dataset, we conduct human validation by 4 dementia domain experts to
verify that the retained dialogues exhibit emotional states, linguistic patterns,
action behaviors, and memory impairments consistent with the corresponding dementia
subtype and disease stage. 
We use this multi-agent framework to generate the first LLM-based dementia dialogue dataset named DemMA-Dialogue. The  statistics are shown in Table \ref{demma-dilogue}.

\par\smallskip
\captionsetup{font=footnotesize, labelfont=bf}
\scriptsize
\setlength{\tabcolsep}{2pt}
\renewcommand{\arraystretch}{0.95}

\begin{tabularx}{\columnwidth}{%
Y r @{\hspace{4pt}\textbar\hspace{4pt}} Y r}
\toprule
\textbf{Motion} & \textbf{\# (Pct.)} &
\textbf{Sound} & \textbf{\# (Pct.)} \\
\midrule
lowering head & 4,380 (27.8\%) &
verbal hesitation (um / uh) & 13,703 (57.6\%) \\
fidgeting & 3,096 (19.6\%) &
sighing & 2,991 (12.6\%) \\
looking around & 2,347 (14.9\%) &
murmuring / self-talk & 2,602 (10.9\%) \\
pushing caregiver away & 686 (4.3\%) &
repetitive words & 2,197 (9.2\%) \\
touching forehead & 590 (3.7\%) &
silence for several seconds & 1,592 (6.7\%) \\
standing up & 488 (3.1\%) &
crying & 399 (1.7\%) \\
others & 4,179 (26.5\%) &
groaning in pain & 313 (1.3\%) \\
\midrule
\textbf{Facial expressions} & \textbf{\# (Pct.)} &
\textbf{Cognitive profile} & \textbf{\# (Pct.)} \\
\midrule
frowning & 5,600 (37.0\%) &
has remote episodic memory & 13,303 (83.0\%) \\
avoiding eye contact & 4,045 (26.8\%) &
has semantic memory & 11,468 (71.6\%) \\
vacant expression & 3,662 (24.2\%) &
benefits from cues & 9,538 (59.5\%) \\
smiling & 857 (5.7\%) &
has recent episodic memory & 9,160 (57.2\%) \\
others & 959 (6.3\%) &
retrieval deficit & 6,328 (39.5\%) \\
\midrule
\multicolumn{4}{l}{\textbf{Corpus statistics}} \\
\multicolumn{4}{l}{2,709 dialogues (avg.\ 5.90 turns; median 6; P25/P75 5--7; min/max 3--8)} \\
\multicolumn{4}{l}{Avg.\ action labels / dialogue: 20.19} \\
\bottomrule
\end{tabularx}
\vspace{-1.5em}

\noindent\captionof{table}{Statistics of DemMA-Dialogue.}\label{demma-dilogue}

\normalsize
\par\medskip

\section{Chain-of-Thought distillation agent training}
\subsection{Reasoning Annotation}

To distill high-quality dialogue generation into a single
model, we construct turn-level \emph{reasoning annotations} as intermediate supervision.
Each annotation makes explicit the reasoning trace underlying a dialogue turn, enabling
the model to reproduce comparable utterance and action decisions at inference time
without executing the full agentic generation pipeline.
For each turn $t$, we generate a reasoning annotation formally,
\begin{equation}
E_t = \operatorname{Reason}\!\left(\mathcal{P}, \hat{M}, C_t;\, U_t, A_t\right),
\end{equation}

Each $E_t$ includes: (i) dialogue-state analysis, (ii) caregiver intent, (iii) memory accessibility,
(iv) emotion inference with ICF-b126 justification, and (v) action decision with clinical rationale.
We use $E_t$ as intermediate supervision to learn clinically consistent reason--speak--act behavior.
A concrete example of the turn-level reasoning annotation is provided in
Appendix~\ref{sec:appendix-planner-annotation-example}.

\subsection{Multi-Task Supervised Fine-Tuning}
To jointly model planner rationales, patient utterances, and multimodal action labels within a single forward pass, we adopt a multi task SFT objective. We optimize a weighted sum of (i) masked next-token prediction losses for the \texttt{[PLAN]} and \texttt{[SPEAK]} segments and (ii) a multi-label action classification loss:
\begin{equation}
\mathcal{L}_{\text{total}} = w_{p}\mathcal{L}_{\text{planner}} + w_{u}\mathcal{L}_{\text{utterance}} + w_{a}\mathcal{L}_{\text{action}}.
\label{eq:mtl_total}
\end{equation}
To reduce interference between reasoning and surface-form generation, we apply segment specific token masking (with index sets $T_p$ and $T_u$) so that planner and utterance supervision is decoupled at the token level:
\begin{equation}
\mathcal{L}_{\text{planner}} = 
- \sum_{t} \mathbf{1}(t \in T_p)\,
\log P_{\theta}\big(y^{p}_t \mid y^{p}_{< t}, x\big),
\label{eq:planner-loss}
\end{equation}
\begin{equation}
\mathcal{L}_{\text{utterance}} = 
- \sum_{t} \mathbf{1}(t \in T_u)\,
\log P_{\theta}\big(y^{u}_t \mid y^{u}_{< t}, x\big).
\label{eq:utterance-loss}
\end{equation}
Action labels are predicted by an auxiliary multi-label head and optimized jointly via $\mathcal{L}_{\text{action}}$, encouraging consistent reasoning-to-action behaviors during generation.

\textbf{Multi-Label Action Learning.}
Action prediction is formulated as a multi-label classification task over behavioral cues such as movement, facial expression, and vocal properties. We use a focal-modulated loss:
\begin{equation}
\begin{aligned}
    L_{\text{action}} = & \sum_{k} \left[ - a_k \log \sigma(z_k) \right. \\
    & \left. - (1 - a_k) \log (1 - \sigma(z_k)) \right] (1 - p_t)^2
\end{aligned}
\end{equation}
where $a_k$ is the ground-truth label and $z_k$ is the predicted logit. The focal term emphasizes low-confidence predictions, improving robustness under sparse action distributions.

\textbf{Integrated Learning.}
By training all components jointly in a single forward pass, the model learns dependencies between memory impairment, reasoning, language generation, and multimodal actions. This integrated optimization enables DemMA to produce coherent reasoning traces, realistic patient dialogue, and aligned action labels within a unified model, without separate modules or post-processing.

\section{Experimental Results and Analysis}
\subsection{Experiment Settings}
We fine-tune DemMA on a dementia dialogue corpus annotated with
planner, utterance, and action labels, using an 85\%/15\% train/validation split
with a fixed random seed.
The base model is Qwen3-8B.
Training is conducted for up to 5 epochs with early stopping 
using AdamW 8bit optimization with a learning rate of $5\times10^{-6}$.
All training and experiments are performed with mixed-precision training on 8 NVIDIA H100 GPUs. The code is available at:  \url{https://anonymous.4open.science/r/DemMA-0A4B/}.


\begin{table*}[t]
\centering
\caption{LLM-based judgments and Expert evaluation on patient simulation fidelity and multi-turn dialogue quality.}
\label{tab:joint_eval}

\scriptsize
\setlength{\tabcolsep}{2.6pt}      
\renewcommand{\arraystretch}{1.05}     
\setlength{\extrarowheight}{-0pt}    
\setlength{\aboverulesep}{0.2ex}       
\setlength{\belowrulesep}{0.2ex}
\setlength{\cmidrulesep}{0.2ex}

\begin{tabular}{l|l|cccccc|ccc}
\toprule
\multirow{2}{*}[-0.6ex]{\textbf{Method}} &
\multirow{2}{*}[-0.6ex]{\textbf{Evaluator}} &
\multicolumn{6}{c|}{\textbf{Patient Simulation Fidelity}} &
\multicolumn{3}{c}{\textbf{Multi-turn Dialogue Quality}} \\
\cmidrule(lr){3-8}\cmidrule(lr){9-11}
& &
\textbf{Auth.} & \textbf{Med.} & \textbf{Mem.} & \textbf{Emo.} & \textbf{Act.} & \textbf{Avg.} &
\textbf{Pers.} & \textbf{Lang.} & \textbf{Avg.} \\
\midrule

\multirow{4}{*}[-0.6ex]{\makecell{Vanilla}}
& GPT-5.2-pro          & 1.50 & 2.39 & 2.28 & 3.17 & 0.72 & 2.01 & 2.16 & 1.50 & 1.83 \\
& Gemini-2.5-pro       & 1.60 & 2.00 & 2.00 & 2.60 & 1.00 & 1.84 & 1.80 & 1.60 & 1.70 \\
& Qwen3-32B            & 1.58 & 2.15 & 1.57 & 2.10 & 0.95 & 1.67 & 2.05 & 2.95 & 2.50 \\
& \textbf{Expert Evaluation}       & 1.50 & 2.00 & 2.00 & 2.50 & 1.00 & 1.80 & 2.00 & 1.50 & 1.75 \\
\midrule

\multirow{4}{*}[-0.6ex]{\makecell{Clinical-Profile\\Prompt}}
& GPT-5.2-pro          & 2.14 & 2.11 & 2.78 & 3.56 & 1.31 & 2.38 & 2.33 & 1.67 & 2.00 \\
& Gemini-2.5-pro       & 2.26 & 2.33 & 1.78 & 2.78 & 1.22 & 2.07 & 2.22 & 1.56 & 1.89 \\
& Qwen3-32B            & 2.21 & 2.30 & 1.65 & 2.25 & 1.35 & 1.95 & 2.30 & 3.15 & 2.73 \\
& \textbf{Expert Evaluation}       & 2.00 & 2.50 & 2.00 & 3.00 & 1.50 & 2.20 & 2.50 & 2.00 & 2.25 \\
\midrule

\multirow{4}{*}[-0.6ex]{\makecell{SFT-Utterance}}
& GPT-5.2-pro          & 2.22 & 2.11 & 2.78 & 4.00 & 3.85 & 2.99 & 2.33 & 2.89 & 2.61 \\
& Gemini-2.5-pro       & 2.33 & 2.44 & 3.11 & 3.33 & 3.78 & 3.00 & 2.31 & 3.22 & 2.76 \\
& Qwen3-32B            & 2.30 & 2.35 & 3.10 & 3.00 & 3.75 & 2.90 & 2.40 & 3.65 & 3.03 \\
& \textbf{Expert Evaluation}       & 2.50 & 2.50 & 3.00 & 3.50 & 4.00 & 3.10 & 2.50 & 3.00 & 2.75 \\
\midrule

\multirow{4}{*}[-0.6ex]{DemMA}
& GPT-5.2-pro          & 3.78 & 4.33 & 4.44 & 4.89 & 4.00 & 4.29 & 4.11 & 3.78 & 3.95 \\
& Gemini-2.5-pro       & 4.06 & 4.44 & 4.12 & 4.75 & 3.75 & 4.22 & 4.44 & 4.56 & 4.50 \\
& Qwen3-32B            & 4.00 & 4.42 & 4.25 & 4.15 & 3.95 & 4.15 & 4.35 & 4.75 & 4.55 \\
& \textbf{Expert Evaluation}       & 3.50 & 4.00 & 3.50 & 4.00 & 3.50 & 3.70 & 4.00 & 3.50 & 3.75 \\
\bottomrule
\end{tabular}
\end{table*}

\subsection{Evaluation Metrics}
We evaluate generated dialogues across seven complementary dimensions. Each dimension metric is assessed by LLM evaluators, clinical experts,  students, enabling a multi-perspective evaluation spanning medical, linguistic, and behavioral criteria.

\textbf{Personality Consistency.}
Measures whether the agent maintains a stable and coherent personality across multi-turn dialogue, including consistent identity, preferences and behavioral tendencies.

\textbf{Language Naturalness.}
Measures whether utterances sound spontaneous and spoken-like rather than templated or model-polished.

\textbf{Authenticity.}
Measures how realistic the dialogue is, as natural fluencies, repetition, and variability.

\textbf{Medical Consistency.}
Measures alignment of language and behavior affect with the specified dementia subtype and severity.

\textbf{Memory Rationality.}
Measures whether forgetting, repetition, and cue responses follow the defined memory profile.

\textbf{Emotional Reasonableness.}
Measures whether emotion shifts are context-driven and gradual rather than abrupt or random.

\textbf{Action Alignment.}
Measures whether nonverbal actions are plausible and consistent with the utterance and clinical profile.

\subsubsection{Automated Evaluation}
To leverage LLMs’ capabilities as a judge for evaluating
open-ended tasks, we use state-of-the-art LLMs, in-
cluding GPT-5.2-pro, Gemini-2.5-pro and Qwen3-32B models as evaluators.
We design seven task-specific LLM-judge prompts, each corresponding to one of the evaluation dimensions above. 
Judges return a scalar rating (0--5) and a short justification. All judge prompts are provided in the Appendix~\ref{LLM-Based Judge Prompts}.

\subsubsection{Human Evaluation}
For dataset annotation, we recruited three dementia experts, each with over ten years of clinical experience in dementia diagnosis and patient care. For evaluation, we additionally recruited five experts with comparable clinical backgrounds and four medical students with formal training in neurology or geriatrics.

\subsection{Baselines}
We compare \textsc{DemMA} with three baselines that progressively add dementia-specific conditioning and training signals: (i) \textbf{Vanilla}, using only background profile without dementia cues; (ii) \textbf{Clinical-Profile Prompt}, which adds a pathology-specific persona describing subtype symptoms; (iii) \textbf{SFT-Utterance}, a supervised model trained to generate utterances without planner reasoning or action supervision. 

\section{Results and Analysis}
In this section, we present comprehensive experiments, aiming to address the following Research Questions (RQs):
\noindent\textbf{RQ1: Simulation fidelity.} Can \textsc{DemMA} achieve high-fidelity simulation of dementia patients in different subtypes?

\noindent\textbf{RQ2: Dialogue quality.} Can \textsc{DemMA} sustain high-quality multi-turn interactions, maintaining persona consistency and natural conversational interactions?

\noindent\textbf{RQ3: Validity of LLM-based evaluation.} Can LLM-based judges reliably evaluate dementia simulation quality? Do their ratings align with assessments from human experts?

\noindent\textbf{RQ4:Dataset quality and diversity.} Does DemMA-Dialogue provide sufficient diversity and coverage for effective multi-turn agent training?

\noindent\textbf{RQ5: Educational Effectiveness.} Can \textsc{DemMA} support medical training (e.g., medical students or trainee caregivers) by helping them better understand dementia clinical manifestations?

\subsection{Simulation Results}
We evaluated each method using LLM judge in all dementia personas using the same seven metrics. To obtain a single overall score for a method, we average metric ratings within each persona across the nine personas. The automated and human evaluation results are shown in Table \ref{tab:joint_eval}.

To answer \textbf{RQ1}, we focus on authenticity, medical consistency, memory rationality, emotional reasonableness and action alignment evaluations.
\textbf{DemMA consistently achieves the highest overall patient simulation fidelity across all evaluators.}
Across GPT-5.2, Gemini-2.5, and Qwen3-32B judges, DemMA attains the top average scores (4.1–4.3), substantially outperforming all baselines. This consistency across evaluators indicates that DemMA’s gains reflect a consensus improvement rather than evaluator-specific bias.
\textbf{DemMA’s improvement is comprehensive, spanning all key simulation dimensions.}
DemMA achieves the highest or tied-highest scores in all simulation fidelity metrics without exhibiting a clear weakness. In contrast, other methods show pronounced imbalances, performing adequately on language while underperforming on clinically grounded or behavioral dimensions.
\textbf{The progression from prompt-based methods to SFT and ultimately to DemMA reflects a qualitative shift rather than a linear improvement.}
Other approaches remain around score of 2.0 to 3.0, while  DemMA surpasses the 4.0 threshold simultaneously across all evaluation dimensions. This suggests that DemMA introduces a fundamentally different level of modeling capacity, enabling a transition from surface-level conversational plausibility to high-fidelity, clinically grounded patient simulation.

To answer \textbf{RQ2}, we focus on persona consistency and language naturalness evaluations. 
\textbf{DemMA consistently achieves the highest average multi-turn dialogue quality across all evaluators.}
This result indicates that DemMA is not merely marginally usable in extended interactions, but instead reliably operates in a high-quality regime that approaches human-level performance in multi-turn dialogue.
\textbf{DemMA exhibits a substantial advantage in persona consistency, demonstrating stronger cross-turn stability than all baselines.}
In the Persona Consistency (Pers.) dimension, DemMA significantly outperforms competing methods, suggesting that it more effectively maintains stable persona attributes and behavioral patterns across dialogue turns.
\textbf{DemMA maintains persona consistency without sacrificing language naturalness.} This finding indicates that DemMA does not rely on rigid or templated responses to preserve consistency, but instead sustains long-term persona stability within fluent and natural conversational exchanges.

\subsection{Human Evaluation and LLM judgments for Dementia Subtypes}

\begin{figure}[]
    \centering
    \vspace{-1em}
    \includegraphics[width=0.8\columnwidth]{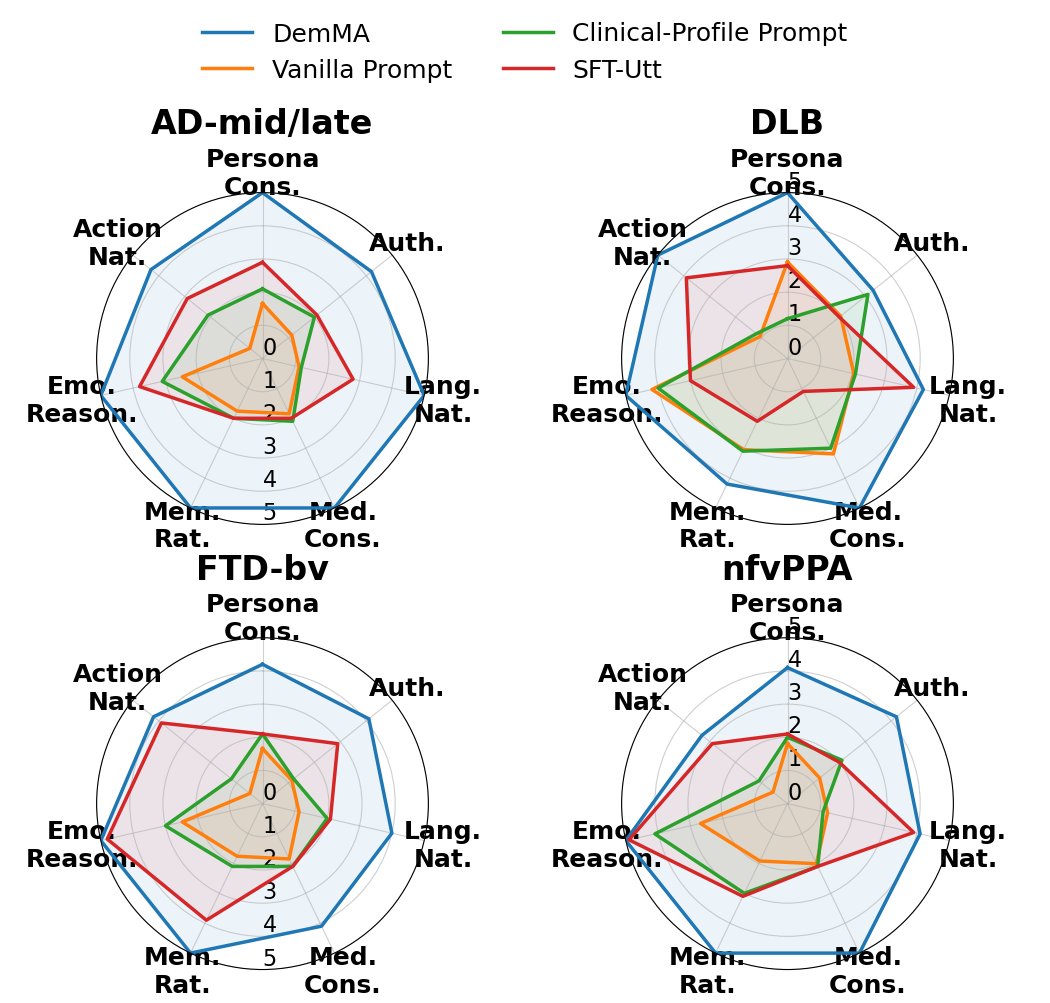}
    \caption{LLM judgment (GPT-5.2-pro) Performance across four dementia subtypes.}
    \label{fig:radar}
    \vspace{-1em}
\end{figure}

To answer \textbf{RQ3}, We conclude the human and LLM judgments evaluation for subtypes in Table \ref{tab:subtype_eval} and Figure \ref{fig:radar}. 
 For Patient Simulation Fidelity, LLM judges score DemMA at 4.15–4.29, while experts assign 3.70; for Multi-turn Dialogue Quality, LLM judges give 3.95–4.55 versus 3.75 by experts. This indicates that \textbf{LLM judges are reliable for comparative evaluation but tend to systematically overestimate absolute quality, particularly for stronger models.} 

 Figure \ref{fig:radar} and Table \ref{tab:subtype_eval} show consistent trends but different emphases. Both indicate that DemMA outperforms baselines across subtypes. However, LLM radar plots amplify performance gaps, while experts provide more conservative, compressed scores (overall 3.5–3.9), and consistently identify action alignment as the main remaining weakness. \textbf{LLM judges are effective for visualizing relative improvements and subtype profiles but expert ratings offer a clinically grounded  assessment.}

\begin{table}[t]
\centering
\caption{Expert evaluation across dementia subtypes (mean of 5 raters).}
\label{tab:subtype_eval}
\scriptsize
\setlength{\tabcolsep}{2.5pt}
\renewcommand{\arraystretch}{1.05}

\begin{tabular}{l|cccccccc}
\toprule
 & \textbf{Auth.} & \textbf{Med.} & \textbf{Mem.} & \textbf{Emo.} & \textbf{Act.} & \textbf{Avg.} & \textbf{Pers.} & \textbf{Lang.} \\
\midrule
AD-early     & 3.0 & 3.2 & 3.4 & 3.8 & 3.0 & 3.4 & 3.8 & 3.6 \\
AD-mid/late  & 3.0 & 3.6 & 3.6 & 3.4 & 4.0 & 3.7 & 4.0 & 4.0 \\
DLB          & 3.6 & 4.4 & 3.4 & 4.0 & 3.2 & 3.7 & 4.2 & 3.0 \\
PDD          & 3.4 & 3.2 & 3.4 & 3.8 & 3.0 & 3.5 & 3.6 & 3.6 \\
VaD          & 3.8 & 4.2 & 3.4 & 4.4 & 3.6 & 3.9 & 4.2 & 3.4 \\
FTD-bv       & 3.8 & 3.2 & 3.4 & 4.0 & 3.0 & 3.6 & 3.8 & 3.6 \\
nfvPPA       & 4.0 & 4.2 & 4.4 & 4.0 & 3.0 & 3.9 & 4.2 & 4.0 \\
lvPPA        & 3.4 & 3.8 & 4.0 & 3.6 & 3.6 & 3.7 & 4.0 & 3.6 \\
svPPA        & 3.6 & 3.8 & 3.4 & 3.2 & 3.6 & 3.7 & 4.6 & 3.0 \\
\midrule
\textbf{Average} 
             & \textbf{3.6} & \textbf{3.7} & \textbf{3.6} & \textbf{3.8} & \textbf{3.3} & \textbf{3.7} & \textbf{4.0} & \textbf{3.5} \\
\bottomrule
\end{tabular}
\end{table}





%

\subsection{Scaling Effect of training dataset}
\begin{figure}[htbp]
    \centering
    \includegraphics[width=\columnwidth]{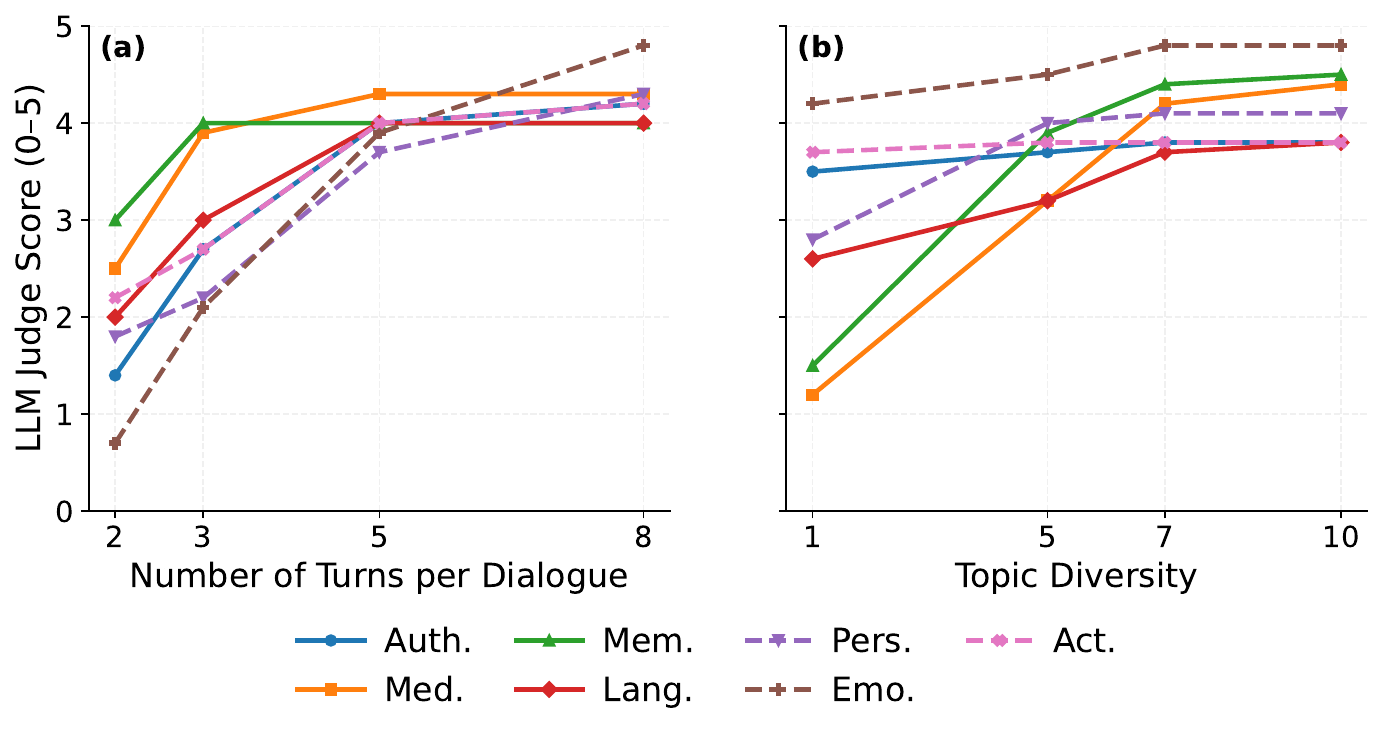}
    \vspace{-1.5em}
    \caption{Scaling Effect of training dataset.}
    \label{fig:Scaling Effect}
    \vspace{-1.5em}
\end{figure}

To answer \textbf{RQ4}, we trained Qwen3-8B model using different scale dataset. Performance is illustrated in Figure \ref{fig:Scaling Effect}.
\textbf{Longer dialogues mainly improve multi-turn capability, but the gains quickly show diminishing returns.}
Increasing the number of turns per dialogue: The largest improvement occurs when moving from short to medium-length dialogues, after which performance tends to plateau. Longer contexts particularly boost Emotional Reasonableness and also stabilize persona consistency, language naturalness, and authenticity, since these dimensions depend on cross-turn coherence and gradual evolution.
\textbf{Increasing topic diversity strongly improves Medical Consistency and Memory  Rationality.}
The model must repeatedly observe how the same pathology manifests across different situations. In contrast, Action Alignment and Authenticity change relatively little, suggesting they rely more on action-label supervision/alignment quality or stylistic cues than on topic coverage alone.

\subsection{Educational effectiveness and feedback}
\begin{figure}[htbp]
    \centering
    \includegraphics[width=\columnwidth]{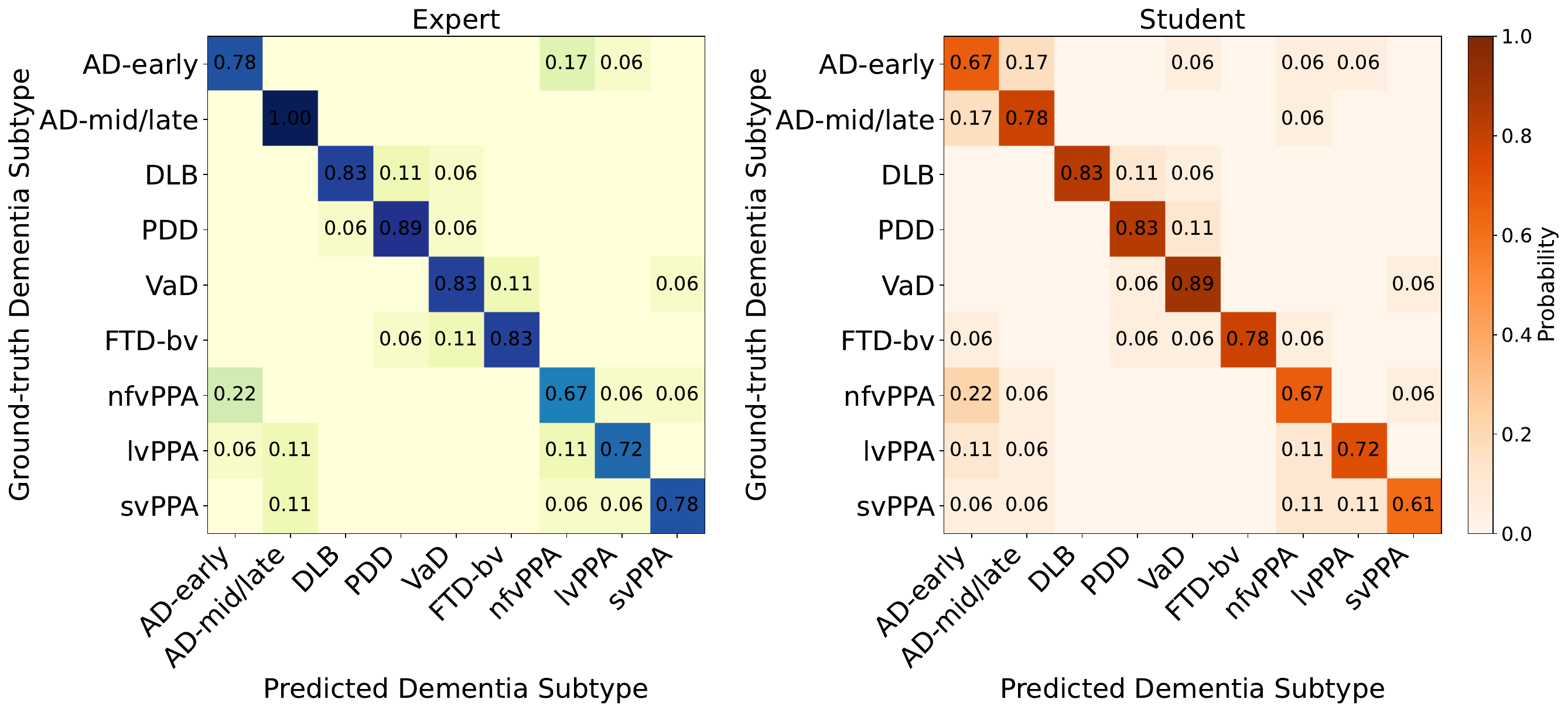}
    \vspace{-1.5em}
    \caption{Confusion matrices for dementia subtype identification (experts vs. students) across nine subtypes.}
    \label{fig:confusion subtype}
    \vspace{-1.5em}
\end{figure}

\begin{figure}[htbp]
  \centering
  \includegraphics[width=\columnwidth]{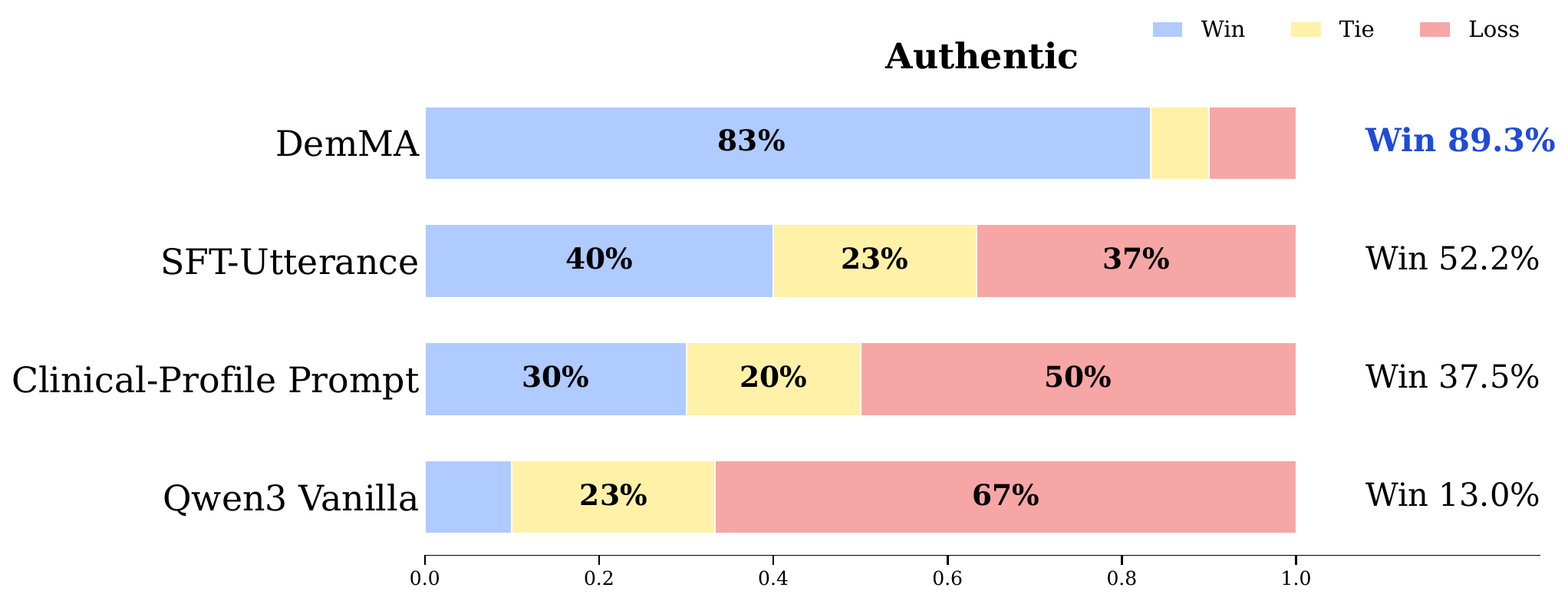}
  \caption{Pairwise win/tie/loss on \textsc{Authentic}. Win rate is computed as $\mathrm{Win}/(\mathrm{Win}+\mathrm{Loss})$, excluding ties.}
  \label{fig:authentic-winrate}
  \vspace{-1em}
\end{figure}

To answer \textbf{RQ5}, we conducted a blinded subtype identification study with medical students and dementia experts. Participants were shown randomized DemMA-generated multi-turn dialogues and asked to predict the dementia subtype for each case. Results are summarized as confusion matrices in Fig.\ref{fig:confusion subtype}.
\textbf{Students can recognize dementia subtype-specific clinical differences via DemMA.}
Both matrices exhibit a strong diagonal, showing that medical students can correctly identify most dementia subtypes from DemMA dialogues, while their accuracy is generally lower than experts.  DemMA provides distinct, learnable clinical cues that differentiate subtypes.
\textbf{Student errors mirror expert confusion in clinically overlapping subtypes.}
DemMA can separate subtypes when cues are clear and reproduces realistic boundary cases.
The pairwise evaluation in Figure \ref{fig:authentic-winrate} shows that DemMA is overwhelmingly preferred for authenticity. DemMA achieves an 89.3\% win rate, demonstrating a clear and robust advantage in perceived dementia realism.


\section{Related Work}
\label{sec:related}

\subsection{CoT Distillation Training}
Chain-of-Thought (CoT) distillation has been proposed as an effective way to retain multi-step reasoning capabilities of large language models while reducing inference-time overhead. Early studies show that exposing models to explicit reasoning traces improves performance on complex tasks, but directly generating such traces during inference is computationally expensive~\cite{wei2022chain}. To address this limitation, subsequent work distills teacher-generated rationales or planning traces into student models via supervised training, enabling reasoning behaviors to be implicitly internalized without explicit CoT generation at inference~\cite{ho2022rethinking,magister2023teaching}. This idea has further been extended to planner–executor and agent-based frameworks, where multi-step decision-making processes are distilled into a single model to maintain long-horizon coherence with reduced latency~\cite{nye2021show,yao2022react}. In contrast to prior CoT distillation methods that primarily focus on textual reasoning, our work distills clinically grounded reasoning signals, including memory analysis, dialogue planning, and action decisions, into a unified model for efficient dementia dialogue generation.

\subsection{LLM-based Patient Simulation for Clinical Training}
LLMs have been explored for dementia-oriented conversational support and caregiver assistance, with recurring concerns around safety and ethics \citep{treder2024llms}. Recent simulated-patient systems such as PATIENT-$\Psi$ and PatientSim ground LLM behavior in structured personas and psychological models to produce plausible multi-turn clinical interactions \citep{wang2024patientpsi,kyung2025patientsim}. However, these frameworks largely assume cognitively stable patients and optimize for persona consistency, which under-represents dementia-specific phenomena such as repetition, contradictory recall, and temporal disorientation~\citep{Lyketsos2011ADNPS,Zhao2016MetaNPS,Aalten2008EADCNPSPartII}. In contrast, DemMA models dementia as a progressing cognitive state and generates clinically grounded breakdown patterns that evolve across turns.


\label{sec:related2}

\section{Conclusion}

In this paper, we presented DemMA, a clinically grounded multi-turn dementia dialogue agent that simulates realistic behaviors through joint modeling of language and action generation. DemMA integrates persona-driven dementia subtypes and disease stages with a structured pipeline, enabling controlled simulation of cognitive decline, emotional variation, and nonverbal cues. We introduce a distillation-style strategy that internalizes planning, reasoning, dialogue, and action prediction into a single low-latency model. Extensive evaluations show that DemMA outperforms baselines in persona fidelity, clinical validity, and educational effectiveness, offering a scalable and ethically sound solution for high-fidelity simulation avoiding sensitive real-world data.
\section*{Limitations}

Despite strong results, DemMA has several limitations.
First, the dataset is fully synthetic. Although DemMA-Dialogue avoids sensitive patient data and is validated by domain experts, synthetic interactions may omit rare clinical behaviors and idiosyncratic caregiver–patient dynamics, potentially limiting generalization to real-world settings. Second, clinical grounding relies on textual abstractions. Our action labels serve as interpretable proxies for multimodal behaviors but cannot fully capture fine-grained audiovisual or sensorimotor cues, constraining fidelity for embodied or multimodal applications. Third, persona and memory modeling are simplified. While we cover multiple dementia subtypes and memory accessibility patterns, the framework does not explicitly model long-term disease progression, medication effects, or evolving caregiver strategies. Finally, evaluation depends partly on LLM-based judges. Such judges may exhibit known biases and may not fully capture clinically grounded correctness \citep{yao2022react}. We therefore treat automated scores as complementary to expert and human evaluation rather than definitive clinical assessments.

\section*{Ethical Considerations}

DemMA is designed to address ethical and legal barriers in dementia research by generating fully synthetic dialogue data rather than relying on real patient records. This substantially reduces risks related to privacy, consent, and exposure of sensitive health information. Nevertheless, synthetic data may still resemble real individuals or scenarios, and dataset release should include screening for personal identifiers and clear usage guidelines.

The framework is intended for research and educational purposes, such as caregiver training and model development, and is not suitable for clinical diagnosis or decision-making. Without appropriate guardrails, simulated dementia dialogue could be misused to generate persuasive but medically incorrect content. Any downstream deployment should therefore include explicit scope restrictions, disclaimers, and human oversight. Simulating cognitive impairment also raises concerns about representational harms and stereotyping. Although personas are clinically informed and validated, model outputs may oversimplify or bias portrayals of dementia. We encourage future audits across demographic and interactional dimensions to assess and mitigate such risks.
Finally, automated evaluation relies partly on LLM-based judges, which are known to exhibit biases and limitations. For this reason, we complement automated metrics with expert and human evaluations and emphasize that reported scores should not be interpreted as clinical assessments.

\bibliography{custom}

\newpage

\end{document}